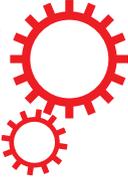

# OPEN  One node driving synchronisation

Chengwei Wang, Celso Grebogi & Murilo S. Baptista

Abrupt changes of behaviour in complex networks can be triggered by a single node. This work describes the dynamical fundamentals of how the behaviour of one node affects the whole network formed by coupled phase-oscillators with heterogeneous coupling strengths. The synchronisation of phase-oscillators is independent of the distribution of the natural frequencies, weakly depends on the network size, but highly depends on only one key oscillator whose ratio between its natural frequency in a rotating frame and its coupling strength is maximum. This result is based on a novel method to calculate the critical coupling strength with which the phase-oscillators emerge into frequency synchronisation. In addition, we put forward an analytical method to approximately calculate the phase-angles for the synchronous oscillators.

A remarkable phenomenon in phase-oscillator networks is the emergence of collective synchronous behaviour[1–6] such as phase synchronisation or phase-locking[7–11]. The Kuramoto model[12–14], a paradigmatic network to understand behaviour in complex networks, has drawn lots of attention of scientists[15–19]. Many incipient works about Kuramoto model have assumed an infinite amount of oscillators coupled by a homogeneous strength. In 2000, Strogatz wrote[20]: "As of March 2000, there are no rigorous convergence results about the finite-N behavior of the Kuramoto model." Since then, understanding the behaviour of networks composed by a finite number of oscillators[21–28] coupled by heterogeneously strengths[29,30] has been the goal of many recent works towards the creation of a more realistic paradigmatic model for the emergence of collective behaviour in complex networks.

However, most of the works about the finite-size Kuramoto model have relied on a mean field analysis, and consequently the emergence of synchronous behaviour has been associated with the collective action of all oscillators. Little is known about the contribution of an individual oscillator into the emergence of synchronous behaviour. But emergent behaviour in real complex networks can be tripped by only one node. Understanding the mechanism behind such a phenomenon in a paradigmatic, more realistic phase-oscillator network model is a fundamental step to develop strategies to control behaviour in complex systems. Besides, no analytical work has been proposed to solve the phase-angles of the synchronous oscillators. But a solution for the phase-angles is of great importance as, for example, they are key variables for monitoring generators in the power grids where a Kuramoto-like model is considered[31–33].

In this paper, we firstly provide a novel method to calculate the critical coupling strength that induces synchronisation in the finite-size Kuramoto model with heterogeneous coupling strengths. From our theory, we understand that the synchronisation of a finite number of oscillators is surprisingly independent of the distribution of their natural frequencies, weakly depends on the network size, but remarkably depends on only one key oscillator, the one maximising the ratio between its natural frequency in a rotating frame and its coupling strength. This lights a beacon for us that in order to predict, enhance or avoid synchronisation in a network of arbitrary size, all required is the knowledge of the state of only one node rather than the whole system. Under a practical point of view, if a pinning control[34,35] would be applied to enhance or slack synchrony in the studied network, the control function can be input into only one node. In addition, we put forward an analytical method to approximately calculate the phase-angles of synchronous oscillators, without imposing any restriction on the distribution of natural frequencies. This directly links the synchronous solution and the physical parameters in phase-oscillator networks.

## Results

**Software codes.** All the software codes for this paper are available by searching at http://pure.abdn.ac.uk:8080/portal/

**Critical coupling strength.** We use $\vec{1}_N (\vec{0}_N)$ to denote the $N \times 1$ vector with all elements equal to one (zero), $\mathbb{I}_N$ to indicate the index set $\{1, 2, \cdots, N\}$. Given a vector $\vec{a}$ with $N$ elements, we use $\overline{a} = \frac{1}{N}\sum_{i=1}^{N} a_i$ to denote the mean value of the elements of $\vec{a}$. The finite-size Kuramoto model with heterogeneous coupling strengths for all-to-all networks is defined as,

Institute for Complex Systems and Mathematical Biology, King's College, University of Aberdeen, Aberdeen, AB24 3UE, UK. Correspondence and requests for materials should be addressed to C.W. (email: r01cw13@abdn.ac.uk)

$$\dot{\Theta}_i = \Omega_i + \frac{\alpha_i K}{N}\sum_{j=1}^{N} \sin(\Theta_j - \Theta_i), \quad \forall\, i \in \mathbb{I}_N, \tag{1}$$

where $N > 0$ is a finite integer number, $K > 0$ is the coupling strength, $\vec{\Omega} = [\Omega_1, \cdots, \Omega_N]^T, \vec{\Theta} = [\Theta_1, \cdots, \Theta_N]^T$, and $\vec{\alpha} = [\alpha_1, \cdots, \alpha_N]^T$ ($\alpha_i > 0, \forall\, i \in \mathbb{I}_N$), denote the vectors whose elements represent the oscillators' natural frequencies, instantaneous phases, and coupling weights, respectively. Define the frequency synchronisation (FS), i.e., the phase-locking state, of the phase-oscillators described by Eq. (1) as,

$$\dot{\Theta}_i - \dot{\Theta}_j = 0 \text{ as } t \to \infty, \quad \forall\, i,\, j \in \mathbb{I}_N. \tag{2}$$

Our goal is to find $K_C$, as the oscillators emerge into FS for a large enough $K$ with as $K > K_C$.

Let $\nu = \dot{\Theta}_i, \forall\, i \in \mathbb{I}_N$, indicate the instantaneous frequency of the oscillators when FS is reached. Divide by $\alpha_i$ on both sides of Eq. (1), then sum the equation from $i = 1$ to $N$, this results in $\nu = \left(\sum_{i=1}^{N}\frac{\Omega_i}{\alpha_i}\right)/\left(\sum_{i=1}^{N}\frac{1}{\alpha_i}\right)$. We rewrite Eq. (1) in a rotating frame, namely, let $\theta_i \equiv \Theta_i - \nu t$ and $\omega_i \equiv \Omega_i - \nu, \forall\, i \in \mathbb{I}_N$, such that $\dot{\vec{\theta}} = \vec{0}_N$ as the oscillators emerge into FS, and we have,

$$\dot{\theta}_i = \omega_i + \frac{\alpha_i K}{N}\sum_{j=1}^{N} \sin(\theta_j - \theta_i), \quad \forall\, i \in \mathbb{I}_N. \tag{3}$$

Define the order parameter[12,13] by,

$$re^{i\psi} = \frac{1}{N}\sum_{j=1}^{N} e^{i\theta_j}, \quad j \in \mathbb{I}_N, \tag{4}$$

Multiplying $e^{-i\psi}$ on both sides of Eq. (4) and then equating its real part and imaginary part, respectively, we have

$$r = \frac{1}{N}\sum_{j=1}^{N} \cos(\theta_j - \psi) = \frac{1}{N}\sum_{j=1}^{N} \cos\phi_j, \tag{5}$$

$$0 = \sum_{j=0}^{N} \sin(\theta_j - \psi) = \sum_{j=0}^{N} \sin\phi_j. \tag{6}$$

The mean field form of Eq. (3) is $\dot{\theta}_i = \omega_i + \alpha_i Kr \sin(\psi - \theta_i), \forall\, i \in \mathbb{I}_N$. Let $\phi_i = \theta_i - \psi$, and $\zeta_i = \frac{\omega_i}{\alpha_i}$, $\forall\, i \in \mathbb{I}_N$, such that, when FS is reached, i.e., $\dot{\vec{\theta}} = \vec{0}_N$, we have

$$\zeta_i = Kr \sin\phi_i, \quad \forall\, i \in \mathbb{I}_N. \tag{7}$$

Considering $\cos\phi_i = s(i)\sqrt{1 - (\sin\phi_i)^2}$, where $s(i) = \pm 1$, we have, from Eqs. (5) and (7), that,

$$r = \frac{1}{N}\sum_{i=1}^{N} s(i)\sqrt{1 - \left(\frac{\zeta_i}{Kr}\right)^2}. \tag{8}$$

Define a function $f$ as $f(\vec{\theta}) := [f_1(\vec{\theta}), \cdots, f_N(\vec{\theta})]^T$, where $f_i(\vec{\theta}) := \frac{1}{N}\sum_{j=1}^{N}\sin(\theta_j - \theta_i)$, and a set $\mathbb{A}$ as $\mathbb{A} := \{\vec{\theta}: Kf(\vec{\theta}) = -\vec{\zeta}\}$ representing the solution for the synchronisation manifold of Eq. (3). From Eqs. (6) and (7), we know, $\bar{\zeta} = \frac{1}{N}\sum_{i=1}^{N}\zeta_i = 0$. Verwoerd and Mason[26] proved that

$$\mathbb{A} \neq \emptyset \Leftrightarrow \text{Eq. (8) holds with } s(i) = 1, \quad \forall\, i \in \mathbb{I}_N. \tag{9}$$

This conclusion was obtained by a Kuramoto model with a mean field coupling strength, i.e., $\alpha_i = \alpha_j = 1, \forall\, i, j \in \mathbb{I}_N$. However, the conclusion in (9) is still effective for the general case where $\alpha_i \neq \alpha_j$. Because the proof for this conclusion was independent of $\vec{\alpha}$, and the only restriction was $\bar{\zeta} = 0$[26], which is fulfilled when $\alpha_i \neq \alpha_j$. The conclusion in (9) means that if Eq. (3) has at least one FS solution, then Eq. (8) holds with $s(i) = 1, \forall\, i \in \mathbb{I}_N$. This FS solution is obtained for $K \geqslant K_C$, where $K_C$ is the critical coupling strength for FS, which ensures that Eq. (8) holds with $s(i) = 1, \forall\, i \in \mathbb{I}_N$[26]. Our following analysis is under the restriction that $s(i) = 1, \forall\, i \in \mathbb{I}_N$, which implies $\cos\phi_i \geqslant 0$, i.e., $\phi_i \in \left[-\frac{\pi}{2}, \frac{\pi}{2}\right], \forall\, i \in \mathbb{I}_N$.

Define the *key ratio* by,

$$\zeta_m := \frac{\omega_m}{\alpha_m}, \quad m \in \mathbb{I}_N, \quad \text{such that } |\zeta_m| \geqslant |\zeta_i|, \quad \forall\, i \in \mathbb{I}_N, \tag{10}$$

meaning that $\zeta_m$ is the one of $\zeta_i$ possessing the maximum absolute value. We call the $m$-th oscillator as the *key oscillator*. We assume $\zeta_m \neq 0$ by ignoring the particular case where $\zeta_m = 0$ resulting in $\omega_i = 0$ and $\zeta_i = 0, \forall\, i \in \mathbb{I}_N$. Let $x = \sin\phi_m$, where $x \neq 0$ and $\phi_m \neq 0$ obtained from $\zeta_m \neq 0$ and Eq. (7). Then we have, from Eq. (7), that $Kr = \frac{\zeta_m}{x}$. Substituting $Kr = \frac{\zeta_m}{x}$ into Eq. (8), and considering $s(i) = 1, \forall\, i \in \mathbb{I}_N$, $r$ can be calculated by

$$r = \frac{1}{N}\sum_{j=1}^{N}\sqrt{1-\left(\frac{x\zeta_j}{\zeta_m}\right)^2}.$$

(11)

Because $\phi_i \in \left[-\frac{\pi}{2}, \frac{\pi}{2}\right]$ and $|\zeta_m| \geqslant |\zeta_i|, \forall\, i \in \mathbb{I}_N$, we have, from Eq. (7), that $|\sin \phi_m| \geqslant |\sin \phi_i|, \forall\, i \in \mathbb{I}_N$, implying $|\phi_m| \geqslant |\phi_i|, \forall\, i \in \mathbb{I}_N$. Therefore, the $m$-th oscillator (the key oscillator) is the most "outside" one of all FS oscillators spreading on a unit circle, where the most inner oscillator possesses the smallest value of $|\phi_i|$ among all oscillators. As $K$ is decreased from a larger value that enables FS in the network to a smaller one, $|\sin \phi_i|$ (as well as $|\phi_i|$) increases correspondingly since $|\sin \phi_i| = \left|\frac{\zeta_i}{Kr}\right|$ from Eq. (7). For any $i \neq j$, if $|\zeta_i| > |\zeta_j|$, we have $|\sin \phi_i| > |\sin \phi_j|$ from Eq. (7), implying $|\phi_i| > |\phi_j|$. This means that $|\phi_i| > |\phi_j|$ is determined only by the condition $|\zeta_i| > |\zeta_j|$, and is independent of $K$. Thus, if we rank oscillators by their values of $|\phi_i(K)|$, this ranking is not altered as $K$ is varied. This means that, regardless of the value of $K$, the key oscillator is always the most "outside" one. FS stops existing if no solution is found for $\left|\sin \phi_i\right| = \left|\frac{\zeta_i}{Kr}\right|$, for any one oscillator. As $K$ is decreased further, the first oscillator for which $\left|\frac{\zeta_i}{Kr}\right| > 1$ (and therefore, no solution is found for $|\sin \phi_i| = \left|\frac{\zeta_i}{Kr}\right|$) will be the key oscillator, because $|\zeta_m| \geqslant |\zeta_i|, \forall\, i \in \mathbb{I}_N$, such that $\left|\frac{\zeta_m}{Kr}\right|$ exceeds 1 at first. This means that $K_C$ is the smallest $K$ for which the key oscillator has a zero instantaneous frequency in the rotating frame, i.e., $\dot{\theta}_m = 0$, resulting in Eq. (7) as $i = m$ with restrictions $\phi_m \in \left[-\frac{\pi}{2}, \frac{\pi}{2}\right]$ and $\phi_m \neq 0$. Therefore, $K_C$ can be obtained by the following optimisation (OPT) problem in (12) to find the minimum $K$ that implies $K = \frac{\zeta_m}{xr}$ with the restrictions that $x \in [-1, 1]$ and $x \neq 0$, where $r$ is calculated by Eq. (11), namely,

$$\begin{cases} \text{minimize } f(x) = \dfrac{\zeta_m}{\frac{x}{N}\sum_{j=1}^{N}\sqrt{1-\left(\frac{x\zeta_j}{\zeta_m}\right)^2}}, \\ \text{subject to } x_{min} \leqslant x \leqslant x_{max}, \end{cases}$$

(12)

where $x_{min} = \varepsilon^+, x_{max} = 1$ if $\omega_m > 0$, and $x_{min} = -1, x_{max} = \varepsilon^-$ if $\omega_m < 0$, where $\varepsilon^+$ ($\varepsilon^-$) indicates a positive (negative) infinitesimal. OPT in (12) can be numerically solved by selecting a small step for $x$, $x_{step}$, then increasing $x$ from $x_{min}$ to $x_{max}$ by $x_{step}$, such that we get a series of values of $f(x)$. The minimum $f(x)$ is $K_C$.

Explosive synchronisation was studied in ref. 36 using a generalised Kuramoto model, which is a particular case of the model described in Eq. (3) by setting $\alpha_i = |\omega_i|, \forall\, i \in \mathbb{I}_N$. In this case, we have $\zeta_i = \pm 1$, and $\frac{\zeta_m}{x} = Kr > 0$ from Eq. (7). Then OPT in (12) can be analytically solved, and the minimum of $f(x)$ is 2, i.e., $K_C = 2$ when $|x| = \frac{\sqrt{2}}{2}$. This result remarkably coincides with the critical coupling strength proposed in ref. 36 for the backward process (namely, decrease $K$ from a larger one to a smaller one) of the explosive behaviour. However, the critical coupling strength for the backward process is different from the one for the forward process (namely, increase $K$ from a smaller one to a larger one) for the explosive synchronisation[36]. In this paper, we consider network configurations for which the critical coupling strength is the same for both the backward process and the forward process, i.e., no explosive synchronisation happens, then $K_C$ obtained by OPT in (12) is also the critical coupling strength for the onset of FS in the forward process.

We further find, numerically, that OPT in (12) obtains its solution at $|x| \approx 1$. Consider $\frac{\zeta_m}{x} = Kr > 0$, an approximate $K_C$ can be analytically obtained by forcing $|x| = 1$, namely,

$$K_C \approx \frac{|\zeta_m|}{\frac{1}{N}\sum_{j=1}^{N}\sqrt{1-\left(\frac{\zeta_j}{\zeta_m}\right)^2}} = K_A.$$

(13)

Let us now numerically demonstrate the exactness of the OPT in (12) to calculate $K_C$, and Eq. (13) to calculate $K_A$ as the approximation of $K_C$, for different phase-oscillator networks. Let $\delta := \max\{|\dot{\theta}_i - \dot{\theta}_j|\}, \forall\, i, j \in \mathbb{I}_N$, where $\delta = 0$ ($\delta > 0$) indicates that all oscillators (not all oscillators) are in FS. The coupling weight $\alpha_i > 0, \forall\, i \in \mathbb{I}_N$, is generated within[1,10], without losing generality. Figure 1(a–c) show the results for three networks: Fig. 1(a), 10 oscillators with $\vec{\Omega}$ following an exponential distribution; Fig. 1(b), 50 oscillators with $\vec{\Omega}$ following a normal distribution; Fig. 1(c), 100 oscillators with $\vec{\Omega}$ following a uniform distribution. We calculate $K_C$ by OPT in (12), and gradually decrease $K$ from $K = K_C + 0.2$ to $K_C - 0.2$. The results show that if $K > K_C$, $\delta = 0$ with an acceptable error in numerical experiments for all cases, meaning that the oscillators are in FS. If $K < K_C$, $\delta > 0$ implying that the oscillators lose FS for all cases. We note that the oscillators lose FS abruptly at $K = K_C$. This means that our method is effective to calculate $K_C$ for all cases. Figure 1(d–f) demonstrate the effectiveness of Eq. (13) to analytically calculate an approximate $K_C$ by forcing $|x| = 1$. Denote $x_{opt}$ as the value of $x$ that provides $K_C$ by OPT in (12). We define the relative error between 1 and $|x_{opt}|$ as $\eta(x) := \frac{1 - |x_{opt}|}{|x_{opt}|}$, and the relative error between $K_A$ [Eq. (13)] and $K_C$ [OPT in (12)] as $\eta(K_C) := \frac{K_A - K_C}{K_C}$. Figure 1(d–f) show the changes of $\eta(x)$ and $\eta(K_C)$ with respect to $N$ ($N = 3$ to 200), with $\vec{\Omega}$ following exponential, normal and uniform distributions, respectively. The results indicate that $K_A$ is near $K_C$, and $|x_{opt}|$ is close to 1 for all cases. This means Eq. (13) works well to approximately calculate $K_C$ for networks formed by arbitrary number of oscillators with any $\vec{\Omega}$ distributions.

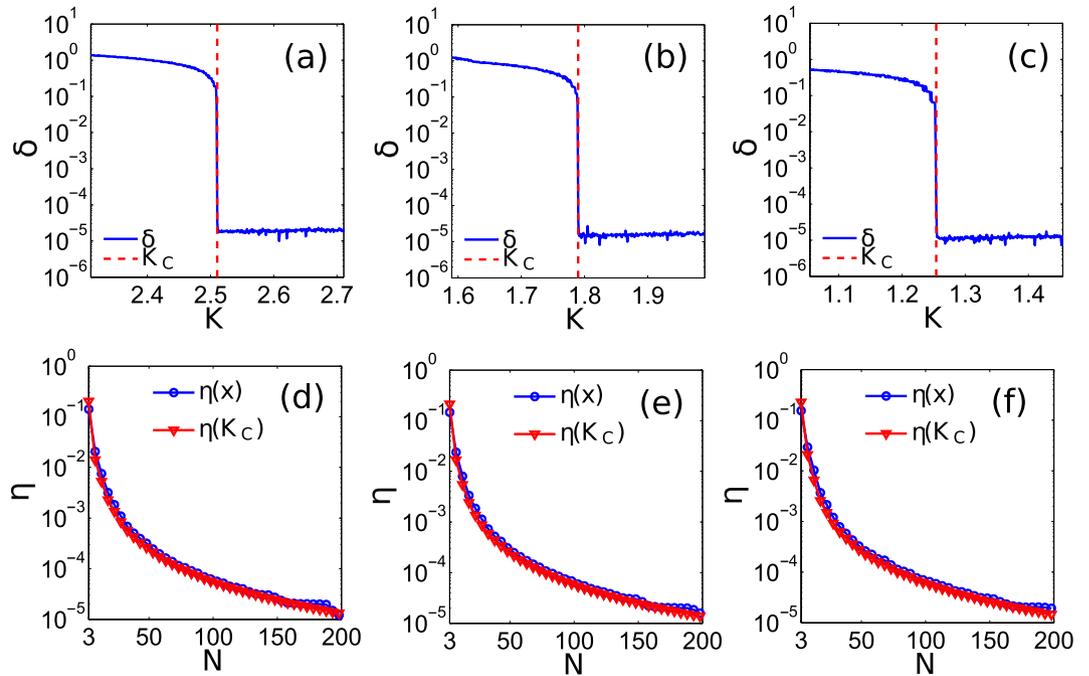

**Figure 1.** (**a–c**) represent the results of $\delta$ (blue solid line) and $K_C$ (red dash line) for networks formed by 10 oscillators with $\vec{\Omega}$ following an exponential distribution, 50 oscillators with $\vec{\Omega}$ following a normal distribution, and 100 oscillators with $\vec{\Omega}$ following an uniform distribution, respectively. (a), (b) and (c) are plotted based on average results of 5000 simulations with different initial phase-angles, but with the same $\vec{\Omega}$ and $\vec{\alpha}$. (**d–f**) show the change of $\eta(x)$ (blue line with circles) and $\eta(K_C)$ (red line with triangles) for networks formed by $N$ ($N = 3$ to 200) oscillators with $\vec{\Omega}$ following exponential, normal and uniform distributions, respectively. (**d–f**) are plotted based on average results of 100 simulations for each $N$, with different $\vec{\Omega}$ and $\vec{\alpha}$.

**One node driving synchronisation.** Below, we show that $K_C$ is independent of the $\vec{\Omega}$ distribution, weakly depends on the network size $N$, and mainly depends only on the key ratio of the key oscillator. For networks with different frequency distributions, diverse network sizes and various key ratios, we verify the dependence of $K_C$ on the $\vec{\Omega}$ distribution, the network size $N$ and the key ratio $\zeta_m$. In order to present the results in a way such that they can be compared, we normalise $\zeta_m$ for these networks by making a parametrisation of $\alpha_m$ based on the value of $\zeta_m$ for each network. The surprising result is that, when we normalise $\zeta_m$ to be the same value for networks with different $N$ and diverse $\vec{\Omega}$ distributions, $K_C$ is roughly the same in these networks. Therefore, the key oscillator is the key factor for the behaviour of these networks. Next, we perform two sets of simulations to demonstrate this result. We use $\vec{\Omega}^e(N)$, $\vec{\Omega}^n(N)$ and $\vec{\Omega}^u(N)$ to denote the natural frequency vectors for networks constructed with a number of $N$ oscillators whose natural frequencies follow exponential, normal and uniform distributions, respectively, and correspondingly use $\zeta_m^e(N)$, $\zeta_m^n(N)$ and $\zeta_m^u(N)$ to indicate the key ratios for these networks.

The first set of simulation includes 6 steps. (i), create all-to-all networks constructed by oscillators with natural frequencies $\vec{\Omega}^e(N)$, $\vec{\Omega}^n(N)$ and $\vec{\Omega}^u(N)$, where $N = 3$ to 200. Thus, we have $3 * (200 - 2) = 594$ networks in total, and each network has a key oscillator with a key ratio $\zeta_m$. (ii), generate the coupling weights for all oscillators in the 594 networks by random numbers in [1, 10]. (iii), find the 594 key oscillators for the 594 networks, and create a set, $\mathcal{O}$, to contain all the 594 key ratios, i.e., $\mathcal{O} := \{\zeta_m^e(N), \zeta_m^n(N), \zeta_m^u(N)\}, \forall N = 3, \cdots, 200$. (iv), find the maximum $\zeta_m$ in $\mathcal{O}$, mark it by $\zeta_s$, and name this key oscillator as the "reference key oscillator" with label $s$. (v), change the values of $\alpha_m$ for all the key oscillators except for the reference key oscillator, such that all $\zeta_m$ are normalised as

$$\left|\frac{\omega_m^e(N)}{\alpha_m^e(N)}\right| = \left|\frac{\omega_m^n(N)}{\alpha_m^n(N)}\right| = \left|\frac{\omega_m^u r(N)}{\alpha_m^u(N)}\right| = \left|\frac{\omega_s}{\alpha_s}\right| = |\zeta_s| = \gamma \zeta_0, \quad \forall N = 3, \cdots, 200,$$

(14)

where $\zeta_0 = |\zeta_s|$ is a constant, and $\gamma$ is a varying parameter which is set to be equal to 1 in the first set of simulation and will vary in the second set of simulation. Note that, this parametrisation process will enlarge all $\zeta_m$ except for $\zeta_s$, such that all of these oscillators maintain their status of key oscillators in their own networks. (vi), calculate and record $K_C$ for all the 594 networks.

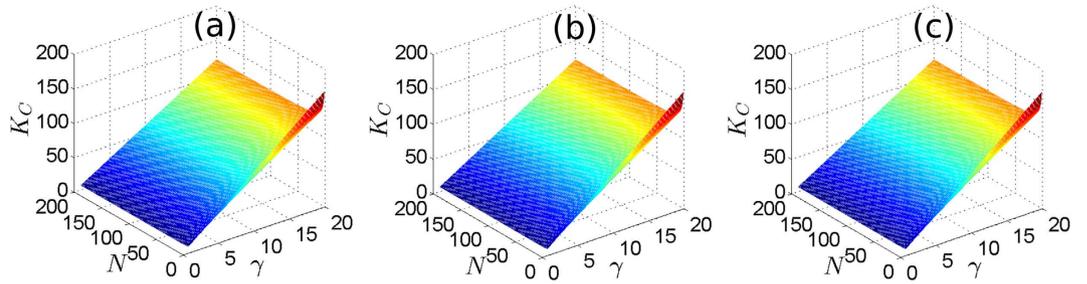

**Figure 2. Exploring the determinant physical parameters for the emergence of the frequency synchronisation.** (**a–c**) show the results for networks formed by $N(N = 3$ to $200)$ oscillators with $\vec{\Omega}$ following exponential, normal, and uniform distributions, respectively. $\gamma$ is the the parameter used to re-scale the key ratio. The surface represents the critical coupling, $K_C$, for different $N$ and $\gamma$.

In the second set of simulation, we further parametrise $\alpha_m$ as a function of $\gamma$ for all the 594 key oscillators. We increase $\gamma$ from its original value 1 to 20 by a small step, and simultaneously decrease each $\alpha_m$ by a proper ratio, such that Eq. (14) still holds. For each value of $\gamma$, we calculate and record $K_C$ for all the 594 networks.

Figure 2(a–c) show the results for networks with frequency vectors given by $\vec{\Omega}^e(N), \vec{\Omega}^n(N)$, and $\vec{\Omega}^u(N)$, respectively. The surfaces representing $K_C$ are similar in all panels, which means that $K_C$ is independent of the $\vec{\Omega}$ distribution. We note that $K_C$ depends on $N$ when $N$ is small, but $K_C$ is almost independent of $N$ for most cases where $N \geqslant 50$. Thus, we say $K_C$ weakly depends on $N$. However, if we keep $N$ unchanged, we observe that $K_C$ almost linearly increases with the growth of $\gamma$ [i.e., the decrease of $\alpha_m^e(N), \alpha_m^n(N)$ and $\alpha_m^u(N)$] for all cases. In other words, $K_C$ will increase if we decrease the coupling weight for only one key oscillator. The reason is that the key oscillator is the first one to lose FS when we decrease $K$, and a key oscillator with a smaller coupling weight is easier to lose FS, which in turn implies a larger $K_C$. As a conclusion, the behaviour of the key oscillator determines the FS of all oscillators, and the key ratio $\left(\zeta_m = \frac{\omega_m}{\alpha_m}\right)$ is the determinant physical parameter for the emergence of FS for all oscillators.

**Master solution.** When the oscillators emerge into FS, i.e., $\dot{\vec{\theta}} = \vec{0}_N$, the solution of Eq. (3) is

$$\vec{\theta} = \vec{\theta}_s + \vec{1}_N \xi, \qquad (15)$$

where $\xi \in \mathbb{R}$ is an arbitrary number, $\vec{1}_N \xi$ is the homogeneous solution of Eq. (3) by setting $\vec{\omega} = \vec{0}_N$, and $\vec{\theta}_s$ is a particular solution of the non-homogeneous Eq. (3). From Eq. (7), we have

$$\phi_i = \arcsin \frac{\zeta_i}{Kr}, \quad \forall \, i \in \mathbb{I}_N, \qquad (16)$$

where we exclude the unstable solutions $\phi_i = \pi - \arcsin \frac{\zeta_i}{Kr}$ for $\zeta_i \geqslant 0$ and $\phi_i = -\pi - \arcsin \frac{\zeta_i}{Kr}$ for $\zeta_i < 0$ (see Methods).

We name $\vec{\phi}$ [Eq. (16)] as the *master solution* of Eq. (3), since it is an analytically expressible particular solution of Eq. (3), and it embodies all of other stable particular solutions, i.e., any stable particular solution $\vec{\theta}_s$ can be expressed by $\vec{\theta}_s = \vec{\phi} + \vec{1}_N \xi$. Note that $r$ in Eq. (16) needs to be numerically calculated. Next, we propose an analytical method to approximately obtain the master solution.

Relabel the oscillators such that $\zeta_1 \geqslant \zeta_2 \cdots \geqslant \zeta_N$, and separate the oscillators into two groups: one group includes oscillators with labels from 1 to $N'$, where $N' = \frac{N}{2}$ (or $\frac{N-1}{2}$) if $N$ is even (or odd); the other group includes the remaining oscillators. Denote $\mu_1 = \frac{1}{N'} \sum_{i=1}^{N'} \zeta_i$ and $\mu_2 = \frac{1}{N-N'} \sum_{i=N'+1}^{N} \zeta_i$ for the first group and second group of oscillators, respectively. From Eqs. (6) and (7), we have $\sum_{i=1}^{N} \zeta_i = 0$. Thus, $\mu_1 N' + \mu_2 (N - N') = 0$, implying $\mu_1 = -\frac{N-N'}{N'} \mu_2 \geqslant 0$. The non-negativity of $\mu_1$ comes from the fact that $\zeta_i \geqslant \zeta_j$ for any $1 \leqslant i \leqslant N'$ and any $N' + 1 \leqslant j \leqslant N$. Recall $N' = \frac{N}{2}$ (or $\frac{N-1}{2}$) if $N$ is even (or odd), we know $N - N' = N'(N - N' \approx N')$ if $N$ is even (odd), implying $\mu_1 = -\mu_2$ ($\mu_1 \approx -\mu_2$) if $N$ is even (odd). For simplicity, we denote $\mu_1 \approx -\mu_2 \geqslant 0$ for both cases. When the oscillators emerge into FS with a given $K'$ ($K' > K_C$), our model treats the whole system as two frequency-synchronous oscillators coupled by a common coupling strength $K'$, with natural frequencies $\mu_1$ and $\mu_2$, respectively. We assume that the two-oscillator system also follows the model described by Eq. (3) with coupling weights $\alpha_1 = \alpha_2 = 1$ which results in $\zeta_1 = \mu_1$, and $\zeta_2 = \mu_2$. Thus, from Eq. (7), we have

$$\begin{cases} \mu_1 = K' r' \sin \phi_1, \\ \mu_2 = K' r' \sin \phi_2, \end{cases} \qquad (17)$$

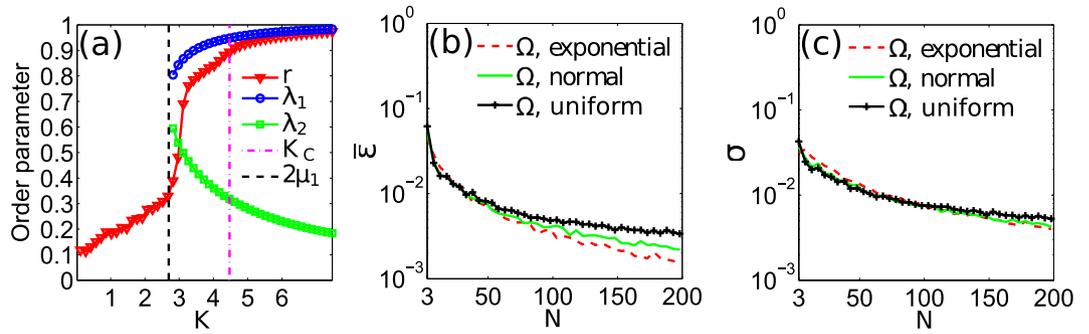

**Figure 3.** (**a**) The order parameter and its approximation for 50 oscillators. $r$ (red line with triangles) is numerically calculated by Eq. (5) as $s(i) = 1, \forall i \in \mathbb{I}_N$. $\lambda_1$ (blue line with circles) and $\lambda_2$ (green line with squares) are calculated by Eq. (18) as $K \geqslant 2\mu_1$. The value of $K_C$ and $2\mu_1$ are represented by magenta dash-dot line and black dash line respectively. (**b,c**) show, for different networks, the change of the average absolute error ($\bar{\varepsilon}$) between $\vec{\phi}'$ in Eq. (19) and $\vec{\phi}$ in Eq. (16) and the standard deviation ($\sigma$) of $\vec{\varepsilon}$, as a function of $K$, respectively. Networks with $N$ (from 3 to 200) oscillators with $\vec{\Omega}$ following exponential (dash red line), normal (green solid line) and uniform (black line with "+") distributions, respectively.

where $r'$ is the order parameter of the two-oscillator system. From Eq. (7), we have $\cos \phi_i = \sqrt{1 - \sin^2 \phi_i} = \sqrt{1 - [\zeta_i/(Kr)]^2}$, where we exclude the case where $\cos \phi_i = -\sqrt{1 - \sin^2 \phi_i}$ (see Methods). Thus, we have $r = \frac{1}{N}\sum_{j=1}^{N}\sqrt{1 - [\zeta_i/(Kr)]^2}$ from Eq. (5). Since $\mu_1 \approx -\mu_2 \geqslant 0$, we have $r' = \frac{1}{2}\left[\sqrt{1 - \left(\frac{\mu_1}{K'r'}\right)^2} + \sqrt{1 - \left(\frac{\mu_2}{K'r'}\right)^2}\right] \approx \sqrt{1 - \left(\frac{\mu_1}{K'r'}\right)^2}$ whose solution is

$$\begin{cases} r_1' \approx \lambda_1 = \frac{\sqrt{2}}{2}\sqrt{1 + \sqrt{1 - \frac{4\mu_1^2}{K'^2}}}, & K' \geqslant 2\mu_1, \\ r_2' \approx \lambda_2 = \frac{\sqrt{2}}{2}\sqrt{1 - \sqrt{1 - \frac{4\mu_1^2}{K'^2}}}, & K' \geqslant 2\mu_1, \end{cases} \quad (18)$$

where $r_1' \approx \lambda_1$ and $r_2' \approx \lambda_2$ indicate a locally stable branch and a locally unstable branch of the FS solution for the two-oscillator model, respectively (see Methods). We only consider the stable branch ($r_1' \approx \lambda_1$). Furthermore, we use the order parameter of the two-oscillator system to be an approximation of the order parameter [Eq. (5)] of the $N$-oscillator system, i.e., $r \approx r' \approx \lambda_1$. Thus, the analytical approximation ($\vec{\phi}'$) for the master solution ($\vec{\phi}$) in Eq. (16) is,

$$\phi_i' = \arcsin\left(\frac{\zeta_i}{K\lambda_1}\right), \quad \forall i \in \mathbb{I}_N. \quad (19)$$

The corresponding approximate FS solution [Eq. (15)], is

$$\theta_i \approx \arcsin\left(\frac{\zeta_i}{K\lambda_1}\right) + \xi, \quad \forall i \in \mathbb{I}_N. \quad (20)$$

Figure 3(a) shows the numerical results of the order parameter for a network with 50 oscillators where $\vec{\Omega}$ follows a normal distribution and $\alpha_i, \forall i \in \mathbb{I}_N$, is a random number within [1,10]. $K_C$ is indicated by the magenta dash-dot line. When $K \geqslant K_C$, the approximate order parameter, $\lambda_1$ [Eq. (18)] is close to the numerical one, $r$ [Eq. (5)]. This means $\lambda_1$ can effectively approximate $r$. Define an $N \times 1$ vector, $\vec{\varepsilon}$, with elements $\varepsilon_i = |\phi_{i'} - \phi_i|, \forall i \in \mathbb{I}_N$ representing the absolute error between $\phi_i'$ [Eq. (19)] and $\phi_i$ [Eq. (16)]. Define $\sigma = \sqrt{\frac{1}{N}\sum_{i=1}^{N}(\varepsilon_i - \bar{\varepsilon})^2}$ as the standard deviation of $\varepsilon_i \in \vec{\varepsilon}$. Figure 3(b,c) show the results of the average absolute error $\bar{\varepsilon}$ and $\sigma$, respectively, at $K = K_C + 0.1$ which ensures the emergence of FS. Networks are formed by $N(N=3$ to $200)$ oscillators, with $\vec{\Omega}$ following exponential, normal and uniform distributions. $\bar{\varepsilon}$ and $\sigma$ are small for all cases, which means that the error between $\phi_i'$ and $\phi_i$ is small $\forall i \in \mathbb{I}_N$ in all cases. Moreover, the larger $K$ is, the smaller the error between $\lambda_1$ and $r$ is [Fig. 3(a)], which will further imply a smaller error between $\phi_i'$ and $\phi_i, \forall i \in \mathbb{I}_N$. This means our method is effective to solve the phase-angles for oscillators as they emerge into FS, for networks formed by an arbitrary number of oscillators with any $\vec{\Omega}$ distribution.

## Discussion

In this paper, we presented our studies on the synchronisation for a finite-size Kuramoto model with heterogeneous coupling strengths. We provided a novel method to accurately calculate [OPT in (12)] or analytically approximate [Eq. (13)] the critical coupling strength for the onset of synchronisation among oscillators. With this method, we find that the synchronisation of phase-oscillators is independent of the natural frequency distribution of the oscillators, weakly depends on the network size, but highly depends on only one node which has the maximum proportion of its natural frequency to its coupling strength. This helps us to understand the mechanism of "the one affects the whole" in complex networks.

In addition, we put forward a method to approximately calculate the phase-angles for the oscillators when they emerge into synchronisation. With our method, one can easily obtain the solution of phase-angles for frequency-synchronous oscillators, without numerically solving the differential equation.

## Methods

**Excluding the unstable solutions.** The FS solution of Eq. (3), i.e., the solution of Eq. (7) is

$$\begin{cases} \phi_i = \arcsin\frac{\zeta_i}{Kr} \text{ or } \phi_i = \pi - \arcsin\frac{\zeta_i}{Kr}, & \text{if } \zeta_i \geqslant 0, \quad \forall \ i \in \mathbb{I}_N, \\ \phi_i = \arcsin\frac{\zeta_i}{Kr} \text{ or } \phi_i = -\pi - \arcsin\frac{\zeta_i}{Kr}, & \text{if } \zeta_i < 0, \quad \forall \ i \in \mathbb{I}_N. \end{cases} \quad (21)$$

A rigorous analysis for the stability of the FS solutions was given by ref. 28 for a mean filed coupled Kuramoto model, i.e., $\alpha_i = \alpha_j = 1, \forall \ i, j \in \mathbb{I}_N$. From the conclusion of ref. 28, we know that the FS solution of Eq. (3) is locally unstable if at least one $s(i) = -1$ in Eq. (8). In other words, if the FS solution is stable, then $s(i) = 1, \forall \ i \in \mathbb{I}_N$, implying that $\cos \phi_i \geqslant 0$, i.e., $\phi_i \in \left[-\frac{\pi}{2}, \frac{\pi}{2}\right], \forall \ i \in \mathbb{I}_N$. Therefore, we exclude the case that $\cos \phi_i = -\sqrt{1 - \sin^2 \phi_i}$ for the solution of the two-oscillator system in the paper.

However, the stability analysis of the FS solution for the general case where $\alpha_i \neq \alpha_j$ is difficult and is still an open problem. In our numerical experiments, the stable solution we obtained is only the one that $\phi_i = \arcsin\frac{\zeta_i}{Kr} \in \left[-\frac{\pi}{2}, \frac{\pi}{2}\right], \forall \ i \in \mathbb{I}_N$. Thus, we exclude the solutions that for $\zeta_i \geqslant 0$, and that for $\zeta_i < 0$.

**The stability analysis for the two-oscillator system.** The two-oscillator system also follows the Kuramoto model with $\alpha_1 = \alpha_2 = 1$, namely,

$$\begin{cases} \dot{\theta}_1 = \mu_1 + \frac{K}{2} \sin(\theta_2 - \theta_1), \\ \dot{\theta}_2 = \mu_2 + \frac{K}{2} \sin(\theta_1 - \theta_2). \end{cases} \quad (22)$$

Let $\vec{\theta}^{eq}$ be a FS solution of Eq. (22). Let $\vec{\theta} = \vec{\theta}^{eq} + \Delta\vec{\theta}$, where $\Delta\vec{\theta}$ is a small perturbation on $\vec{\theta}^{eq}$. Linearise Eq. (22) around $\vec{\theta}^{eq}$, we have,

$$\begin{bmatrix} \Delta\dot{\theta}_1 \\ \Delta\dot{\theta}_2 \end{bmatrix} = \mathbf{J} \begin{bmatrix} \Delta\theta_1 \\ \Delta\theta_2 \end{bmatrix}, \quad (23)$$

where the Jacobian matrix $\mathbf{J}$ is

$$\mathbf{J} = \begin{bmatrix} -\cos(\theta_2^{eq} - \theta_1^{eq}) & \cos(\theta_2^{eq} - \theta_1^{eq}) \\ \cos(\theta_1^{eq} - \theta_2^{eq}) & -\cos(\theta_1^{eq} - \theta_2^{eq}) \end{bmatrix}. \quad (24)$$

The two eigenvalues of $\mathbf{J}$ are $e_1 = 0$ and $e_2 = -[\cos(\theta_1^{eq} - \theta_2^{eq}) + \cos(\theta_2^{eq} - \theta_1^{eq})]$. If the FS solution is stable, we have $e_2 < 0$ implying $\cos(\theta_1^{eq} - \theta_2^{eq}) > 0$, i.e., $|\theta_1^{eq} - \theta_2^{eq}| < \frac{\pi}{2}$.

We have, from Eq. (18), that $\lambda_1\lambda_2 = \frac{\mu_1}{K} \approx -\frac{\mu_2}{K}$. Substituting this condition into Eq. (17), we get $\lambda_1\lambda_2 = r' \sin \phi_1$ and $\lambda_1\lambda_2 = -r' \sin \phi_2$. If $r' \approx \lambda_2$, we have that $\phi_1 \approx \arcsin(\lambda_1)$ and $\phi_2 \approx -\arcsin(\lambda_1)$. Because $\frac{\sqrt{2}}{2} \leqslant \lambda_1 \leqslant 1$ from Eq. (18), we approximately have $\frac{\pi}{4} \leqslant \phi_1 \leqslant \frac{\pi}{2}$ and $-\frac{\pi}{2} \leqslant \phi_2 \leqslant -\frac{\pi}{4}$, implying $|\phi_1 - \phi_2| \geqslant \frac{\pi}{2}$. If $\lambda_1$ grows larger as $K$ increases from $2\mu_1$, $|\phi_1 - \phi_2|$ will become larger. However, $|\phi_1 - \phi_2| = |\theta_1 - \theta_2| \geqslant \frac{\pi}{2}$ implies instability of the FS solution of the two oscillators. This means that $r' \approx \lambda_2$ describes an unstable FS solution. On the other hand, $r' \approx \lambda_1$ ensures the stability of the FS solution.

## Acknowledgements
C.-W.W. is supported by a studentship funded by the College of Physical Sciences, University of Aberdeen. M.S.B. acknowledges EPSRC grant NO. EP/I032606/1.

## Author Contributions
C.-W.W. has perceived the new phenomenon reported in this manuscript, and has performed the simulation and the analytical calculations. M.S.B. and C.G. has contributed with ideas to better explore the implications of this new phenomenon and to the writing of the paper.

## Additional Information
**Competing financial interests:** The authors declare no competing financial interests.

**How to cite this article**: Wang, C. *et al.* One node driving synchronisation. *Sci. Rep.* **5**, 18091; doi: 10.1038/srep18091 (2015).